# A Data Streaming Process Framework for Autonomous Driving By Edge

Hang Zhao[a], Jie Tang[a]*

[a]School of Computer Science and Engineering, South China University of Technology, Guangzhou, China;

*Email address of corresponding author: cstangjie@scut.edu.cn

**ABSTRACT**

In recent years, with the rapid development of sensing technology and the Internet of Things (IoT), sensors play increasingly important roles in traffic control, medical monitoring, industrial production and etc. They generated high volume of data in a streaming way that often need to be processed in real time. Therefore, streaming data computing technology plays an indispensable role in the real-time processing of sensor data in high throughput but low latency. In view of the above problems, the proposed framework is implemented on top of Spark Streaming, which builds up a gray model based traffic flow monitor, a traffic prediction orientated prediction layer and a fuzzy control based Batch Interval dynamic adjustment layer for Spark Streaming. It could forecast the variation of sensors data arrive rate, make streaming Batch Interval adjustment in advance and implement real-time streaming process by edge. Therefore, it can realize the monitor and prediction of the data flow changes of the autonomous driving vehicle sensor data in geographical coverage of edge computing node area, meanwhile minimize the end-to-end latency but satisfy the application throughput requirements. The experiments show that it can predict short-term traffic with no more than 4% relative error in a whole day. By making batch consuming rate close to data generating rate, it can maintain system stability well even when arrival data rate changes rapidly. The Batch Interval can be converged to a suitable value in two minutes when data arrival rate is doubled. Compared with vanilla version Spark Streaming, where there has serious task accumulation and introduces large delay, it can reduce 35% latency by squeezing Batch Interval when data arrival rate is low; it also can significantly improve system throughput by only at most 25% Batch Interval increase when data arrival rate is high.

# 1. INTRODUCTION

In recent years, autonomous driving technology has attracted extensive attention. In addition to the well-known Waymo, other companies such as Uber, Tesla, and Baidu have launched extensive research. In order to realize the environment perception, autonomous driving vehicles are usually equipped with a variety of sensors, such as cameras, RADAR, LIDAR, GPS, IMUs, and so on, which can generate GB level data per minute (Van Rijmenam Mark, 2013).

In one hand, the on-board computing power in autonomous driving vehicles is difficult to support such huge volume of streaming data. In the other hand, the high time delay generated by accessing remote cloud computing center is indeed a big intolerance for latency sensitive tasks such as autonomous driving. With the operation of a large number of autonomous driving vehicles on the road, it will inevitably cause a huge bandwidth pressure on the uplink of the base station. Compared with the shortcomings of the remote cloud center, the edge data centers deployed near base stations have striking geographical advantages. With the peak rate above 10 Gbps, the millisecond-level latency, and the ultra-high-density link support provided by 5G communication technology (Soldani ,et al., 2013), and IPv6 based vehicle network (Liang, W. et al., 2019), sensor data of autonomous driving vehicles can be easily transferred to the nearby edge data center.

Cloud computing data center possesses tremendous storage and computing power, which is suitable for processing tasks such as HD maps storage, building and large-scale neural network training (Liu, S. et al, 2019). On the contrary, the size of the edge data center is relatively small, which is more fit for handling reasoning tasks of small scale and low latency, such as traffic light recognition, pedestrian and vehicle detection (N. Raviteja, et al., 2020), etc. Meanwhile, the edge data center can also provide better position-aware functions. For instance, it could download local HD maps from the distant cloud data center in advance to provide HD map matching service to automated vehicles in its coverage area. Cloud computing and edge computing fusion can improve the performance of vehicle sensor data processing (Lai, Y et al., 2019; Tang, J et al, 2020).

At present, researchers have conducted many studies on the streaming data processing frameworks which are established in the cloud data center. Spark Streaming is one of streaming data processing framework running on Spark developed by the University of California, Berkeley (Zaharia, Li et al., 2013). However, there is rare research about that in the edge data center. In this paper, based on sensor data generated by automated vehicles, we propose a streaming data processing framework, which has the two following advantages:

(1) Based on the gray model (GM), within the coverage scope of a certain edge node, we implement the traffic flow monitor and prediction for autonomous driving vehicles, so that the system can realize its flexibility by conducting the adjustment for resource utilization strategy according to the variation of data stream.

(2) The fuzzy control method is adopted to dynamically adjust the batch interval of Spark Streaming according to the change of data streams and system workload, which contributes to reducing the delay between end to end on the premise of satisfying the throughput requirement.

At this point, we have introduced the application scenarios and the proposed streaming data processing framework for edge computing in this section. The remainder of this paper is organized as follows: Section 2 introduces the related content about the Spark Streaming data processing framework. We depict the overall architecture and system design in Section 3. We present the detailed implementations of each part in Section 4. Section 5 and 6 are experiments and related work respectively. Finally, Section 7 gives the conclusion.

## 2. BACKGROUND AND MOTIVATION

### 2.1 Background

In recent years, with the rapid development of the Internet of Things (IoT) and sensor technology, oceans of sensor devices keep producing uncountable recording data constantly. These data have features of continuity, volatility, and suddenness, commonly known as streaming data. Dealing with these streaming data usually needs to utilize real-time or near real-time processing methods, which is called streaming data processing.

Spark Streaming and Storm are typical streaming data processing frameworks. These frameworks are generally deployed in the remote cloud data center. With the benefits of the strong computing power and storage capacity of the cloud, we can process the received streaming data in real-time. In this processing pattern, sensor data are firstly transmitted in wireless to the gateway (such as base stations), and then transmitted via optical fiber links to the remote cloud data center for processing.

For the past few years, with the deepening explorations on autonomous driving, the types of vehicular sensors are increasingly abundant, and the data generated per unit time are far more than other sensor devices in the past. Hence, these countless data are far beyond the processing capacity of on-board computers, which must be handled with the assistance of cloud computing. However, such a large scale of data is bound to bring heavy bandwidth pressure on the uplink of gateways and cause huge transmission delay in both the process of data uploading and result retrieving. For latency-sensitive tasks such as autonomous driving, it is unacceptable for the instability and the increase of latency during the data transmission process.

In order to solve the shortcomings of streaming data processing in the cloud computing environment, we propose a streaming data processing method oriented to edge computing. Leveraging the computing and storage capacity of the edge, the processing of streaming data produced by autonomous driving is transferred from the remote cloud data center into the nearby edge data center, which decreases the data transmission delay to a great extent.

### 2.2 Problem Analysis

In the data migration process from the distant cloud to the nearby edge, we notice that the existing streaming data processing frameworks, such as Storm and Spark Streaming, are mainly oriented to cloud computing and fail to fully consider the features of data processing under the edge computing environment. As a consequence, they have common limitations including the low resources utilization rate, no real-time adjustment for system resource allocation according to the data input, etc.

Hence, we consider realizing an edge-oriented streaming data processing framework, which meets the demands of sensor data streams processing for autonomous driving. This framework has the following characteristics:

(1) It can forecast short-term traffic and predict the data arrival rate based on the traffic flow, by which it realizes the allocation adjustment for system resources in advance and provides some extent of elasticity for burst traffic.
(2) It strikes a balance between system throughput and end-to-end latency, which reduces the delay as much as possible to satisfy the requirements of delay-sensitive tasks while ensuring the system throughput.

# 3. SYSTEM DESIGN

## 3.1 System Overview

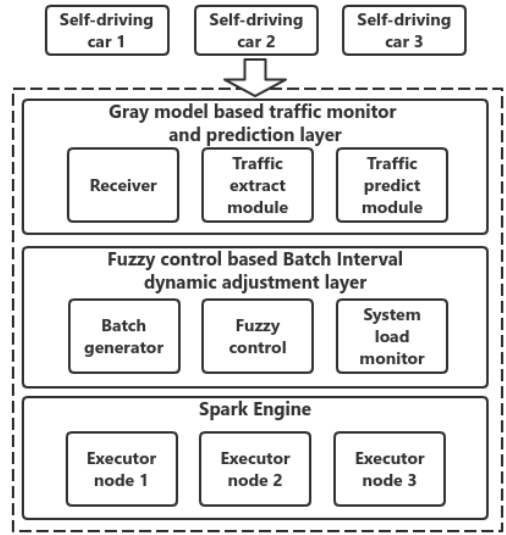

**Figure 1.** System architecture

The proposed framework is deployed in the edge computing node, which takes the data generated by the autonomous vehicles in a specific area as the source data. The system architecture is shown in Figure 1. The proposed framework is designed with three layers: the gray model based traffic monitor and prediction layer, the fuzzy control based Batch Interval dynamic adjustment layer, and the Spark engine executive layer.

## 3.2 The Gray Model Based Traffic Flow Monitor and Prediction Layer

### 3.2.1 Summary

Streaming data not only has a large volume and scale, but also has the characteristics of volume, velocity and verity (3V) (Laney et al., 2001). Therefore, the realization of accurate prediction of streaming data rate can help the streaming big data processing framework to respond better, which provides sufficient computing resources and task scheduling, and enhances the flexibility of the system.

At present, since autonomous vehicles have not been widely used, there is no effective research on the data flow characteristics of autonomous vehicles. We assume that they have the following characteristics:

(1) During the autonomous driving process, each sensor always keeps the working state, and the number and the working mode of sensors remain relatively stable. Then, the same vehicle will continuously generate sensory data during its driving, so that the data rate will remain constant.
(2) For different autonomous vehicles, we assume that the numbers and types of their sensors are similar, and that their data rates are also similar.

In summary, we assume that the total data rate generated by autonomous vehicles is directly proportional to the number of vehicles.

The research shows that the historical traffic flow in the same area has self-similarity (Li et al., 2010). In full consideration of the demand of flow data processing system, we realize the short-term traffic flow prediction based on the gray model, which can model the system with uncertain information through a small amount of sample data, and has the characteristics of small calculation amount and high prediction accuracy. It is widely used in short-term traffic flow prediction (Liu et al., 2006).

In this paper, we implement the pluggable traffic flow monitor and prediction module. By monitoring the flow data received by the Receiver, we can extract the traffic flow data in the current area in real-time. At the same time, the traffic flow monitor and prediction module designed also provides the function of recording the historical data to fit the historical traffic flow data through the gray model and realize the subsequent traffic flow prediction, so as to estimate the data arrival rate required by the flow data processing framework. The flow chart of this module is shown as Figure 2.

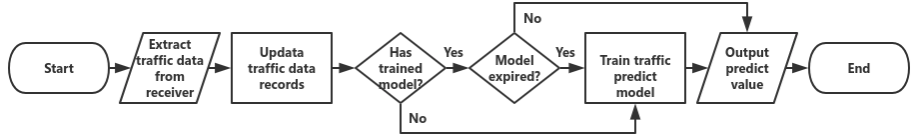

**Figure 2.** The workflow of the traffic monitor and prediction module

### *3.2.2 Gray Model Fitting and Traffic Flow Prediction*

We suppose that the traffic flow monitor module counts the number of vehicles passing in the period T. We record the historical traffic flow data that has been detected as

$$B^{(0)}(t) = [B^{(0)}(1) \quad B^{(0)}(2) \quad \cdots \quad B^{(0)}(n)]^T, \ 1 \leq t \leq n \tag{1}$$

Where n is the number of historical data. For the gray model, it needs a small amount of historical data. We can use only 4-5 data to fit.

Because of the randomness of the original data, it is difficult to use them directly. The gray model can enhance its order by accumulating, so that it can use the exponential differential model to fit. We accumulate the historical traffic data once, and get

$$B^{(1)}(t) = [B^{(1)}(1) \quad B^{(1)}(2) \quad \cdots \quad B^{(1)}(n)]^T, \ 1 \leq t \leq n \tag{2}$$

Where $B^{(1)}(t) = \sum_{i=1}^{t} B^{(0)}(i), \ 1 \leq t \leq n$. (3)

We establish the differential equation as:

$$\frac{dB^{(1)}(t)}{dt} + \alpha B^{(1)}(t) = \mu \tag{4}$$

By the least square method, we get the fitting parameter $\begin{bmatrix}\alpha\\\mu\end{bmatrix}$. Then, taking it into the differential equation, we can get

$$\widehat{B}^{(1)}(t) = \left[B^{(1)}(1) - \frac{\mu}{\alpha}\right] e^{-\alpha(t-1)} + \frac{\mu}{\alpha} \tag{5}$$

By doing the difference, we get the expression of $y_1, y_2, \ldots, y_n, y_{n+1}, \ldots$ as follows:

$$y_t = \begin{cases} \widehat{B}^{(1)}(1), t = 1 \\ \widehat{B}^{(1)}(t) - \widehat{B}^{(1)}(t-1), t > 1 \end{cases} \tag{6}$$

Where $y_1, y_2, \ldots, y_n$ is the fitting result of the passed traffic flow, and $y_{n+1}, y_{n+2}, \ldots$ is the prediction of the future short-term traffic flow.

## *3.3 The Fuzzy Control Based Dynamic Adjustment Layer of Batch Interval*

### *3.3.1 Summary*

Spark Streaming utilizes the D-stream method to form a batch of received data within a certain time (i.e., batch interval), and then the batch is handed over to the bottom Spark engine for processing, so as to achieve high throughput while ensuring real-time performance.

According to research, when the data rate is fixed, the batch interval is closely related to the processing time of the Spark engine (Das T, et al., 2014). If the batch interval is too small, it will cause large scheduling overhead and affect the performance of the Spark engine. If the batch interval is too large, it will increase the end-to-end delay and fail to give full play to its timeliness. In the existing Spark Streaming, once the batch interval is set, it will not change in the life cycle of the application, which cannot adapt to the data rate fluctuation well.

Therefore, we introduce the dynamic adjustment mechanism of batch interval based on fuzzy control. The mechanism adjusts the batch interval according to the change of input data rate and system workload, which reduces the end-to-end delay as much as possible on the premise of meeting the throughput requirement.

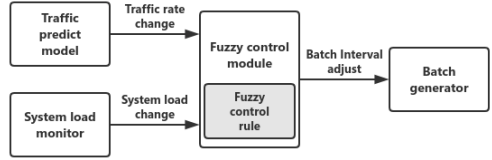

**Figure 3.** The workflow of Batch Interval dynamical adjust

The dynamic adjustment flow of batch interval based on fuzzy control in this paper is shown in Figure 3. The system inputs are the changes of data rate and system workload. The change of data arrival rate comes from the difference between the data arrival rate predicted by the gray model in the next cycle and the real measured data arrival rate in the current cycle, and the change of system processing capacity comes from the runtime data collected by the Spark engine monitor system. The fuzzy control system generates the corresponding output signal by querying the control rule table to control the batch timer to complete the dynamic adjustment of the batch interval.

### *3.3.2 System Workload Monitor*

In order to accurately perceive the changes of system workload and make full use of system resources, we first realize the monitor of system workload. Here, we define the system workload as the ratio of the total end-to-end delay of batch processing in spark streaming to the batch interval.

We assume that there are k tasks running in the system, the system workload detected by task i is defined as

$$\eta_i = \frac{\text{total delay of i-th task}}{\text{batch interval of i-th task}}, 1 \leq i \leq K \quad (7)$$

Because the total delay of the system is greater than 0, it is easy to know that $\eta_i > 0$. According to the definition of $\eta_i$, when $0 < \eta_i \leq 1$, the system can process the task i in a stable state, and process the newly arrived batch in time without any accumulation. In this interval, that $\eta_i$ is too small indicates that the newly arrived batch can be processed quickly and that the system resources are redundant, while that $\eta_i$ is too big indicates that the system resources are fully utilized and that the flexibility of the system is reduced. When the batch data rate increases sharply, the processing delay increases, which is easy to cause task accumulation in the system. When $\eta_i > 1$, the processing capacity of the system cannot keep up with the batch production rate, and the unprocessed tasks in the system will be accumulated, which is easy to cause system crash.

Because of the continuous arrival of streaming data and the continuous processing of batches, the workload monitor of our system should also be a continuous process. Therefore, we select single exponent smoothness model to fit the detected system workload. We make the system workload at time t as $S(t)$, and that at time (t-1) as $S(t-1)$. Then, we can obtain

$$S(t) = \alpha * \frac{\eta_1+\eta_2+\cdots+\eta_k}{K} + (1-\alpha) * S(t-1) \tag{8}$$

Where $\alpha$ is the smoothing coefficient, whose range is (0,1), and $\eta_i, 1 \leq i \leq k$ is the workload information completing batches of the system from time (t-1) to time t.

In this paper, our goal is to ensure that the fitted system workload $S(t) < 1$ to maintain the overall stability of the system and provide some flexibility for the burst data. At the same time, it should be as close as possible to 1, so that each application can reduce the end-to-end delay as much as possible to give full play to the real-time performance of the streaming big data processing framework on the premise of meeting the throughput requirements.

### 3.3.3 Fuzzy Control Rules

Fuzzy control rules are the core of the fuzzy control system. In order to describe fuzzy control rules accurately, we need to define the input, output and rule table of the fuzzy control system.

First of all, we take the rate of change $C(t)$ of traffic flow and the difference $D(t)$ of system workload as the input of the fuzzy system. They are defined as follows:

$$C(t) = \frac{q(t)-q(t-1)}{q(t-1)} - 1 \tag{9}$$

$$D(t) = S(t) - 1 \tag{10}$$

Where $q(t)$ is the data rate at time t and $S(t)$ is the system workload value obtained by the system workload monitor module through single exponential smooth.

We assume that the data range of $C(t)$ and $D(t)$ does not exceed the range of $[-20\%, 20\%]$, and define five states for them respectively: NB (negative big), NS (negative small), ZO (zero), PS (positive small), and PB (positive big). The membership function image of $C(t)$ and $D(t)$ can be obtained as shown in Figure 5. By transforming the specific value into the corresponding state in the membership function, we can achieve fuzzification of the input.

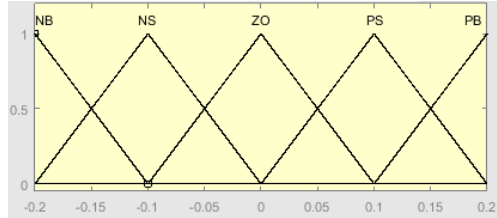

**Figure 4.** C(t) and D(t) have the same Membership Function

Then, we define the value of output f of fuzzy control as $f(-2), f(-1), f(0), f(1), f(2)$, which represent the significant decrease, small decrease, unchanged, small increase and large increase of batch interval respectively.

After defining the input and output, we give the fuzzy control rules as shown in Figure 5. In Figure 5, the change rate $C(t)$ of vehicle flow is represented horizontally, and the difference $D(t)$ of system workload is represented longitudinally. The output of fuzzy control can be determined by the values of $C(t)$ and $D(t)$.

|  |  | Change of traffic: C(t) | | | | |
|---|---|---|---|---|---|---|
|  |  | NB | NS | ZO | PS | PB |
| System workload: D(t) | NB | -2 | -1 | -1 | 0 | 0 |
|  | NS | -1 | -1 | 0 | 0 | 0 |
|  | ZO | -1 | 0 | 0 | 0 | +1 |
|  | PS | 0 | 0 | 0 | +1 | +1 |
|  | PB | 0 | 0 | +1 | +1 | +2 |

**Figure 5.** Expert rule of Fuzzy Control

## 4. SYSTEM IMPLEMENTATION

### *4.1 Overview*

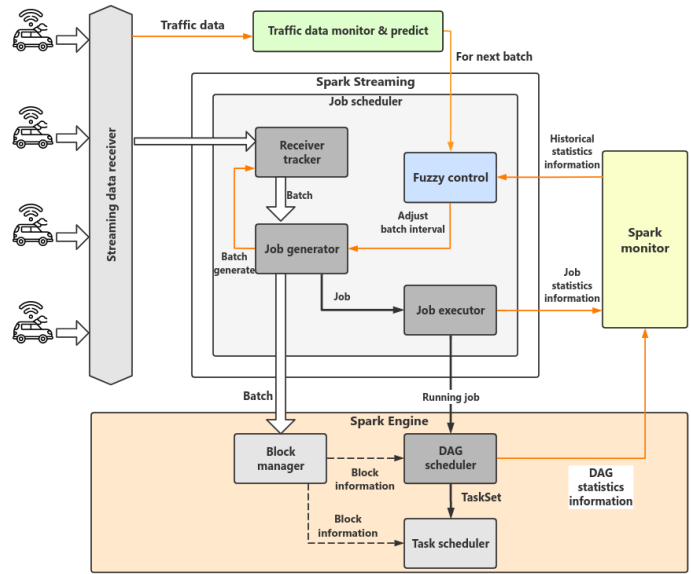

**Figure 6.** System implementation diagram

The edge computing oriented big data processing framework proposed in this paper is implemented on the basis of Spark Streaming. Spark Streaming can complete data reception, partition and DAG generation, then distribute generated DAGs to the underlying Spark engine for distributed execution.

Figure 6 shows the implementation details of Spark Streaming and the modifications we made. Firstly, we add a gray model based traffic flow monitor and prediction layer, which can get the latest traffic data in real time by Receiver and predict the short-term traffic flow. The second section of this chapter will introduce its concrete implementation. Secondly, we implemented the Batch Interval dynamic adjustment layer based on fuzzy control, which can make corresponding adjustment to Batch Interval according to the change of vehicle flow and system workload. The third section of this chapter will introduce its concrete implementation.

### *4.2 The Gray Model Based Traffic Flow Monitor and Prediction Layer*

Figure 7 shows our implementation of the gray model based traffic flow monitor and prediction layer. We implemented the TrafficTracker module to manage historical vehicle traffic data and call gray model to predict short-term traffic flow. At the same time, we modified Receiver implementation so that it can collect traffic information from the received data and upload the extracted traffic to TrafficTracker through RPC.

The class diagram of TrafficTracker, as shown in Figure 8, contains a number of attributes and methods. In the attributes it contains, *trackerState* is responsible for recording the current running state of TrafficTracker,

*endpoint* is a reference to RPC terminal nodes, *timeToTrafficInfos* records the raw data uploaded by Receiver, *resampleInterval* is the resampling cycle, *resampledRecords* is the traffic data after resampling, *trainNum* is the number of samples required for training the gray model, *model* is a well trained model, *lastTrainTime* records the last time of training the gray model. In the method it contains, *start()* and *stop()* are responsible for the start and stop of the module, the *reportInfo()* method is called RPC to add new traffic data, *getInfo()* returns the traffic data of the specified time, the *cleanup()* method is responsible for clearing the old invalid data, the *resample()* method is responsible for resampling the received traffic data, the *getRecords()* method returns the data after resampling, the *getLatestRecord()* method returns the latest one in the resampling data, the *train()* method is responsible for training the gray model, the *getModel()* method returns the training model, and the *predict()* method calls the gray model to predict the traffic volume in the future specified time.

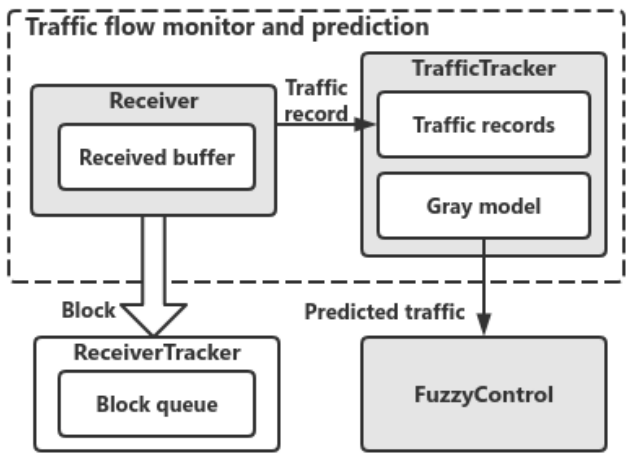

**Figure 7.** Realization diagram of traffic flow monitor and prediction layer

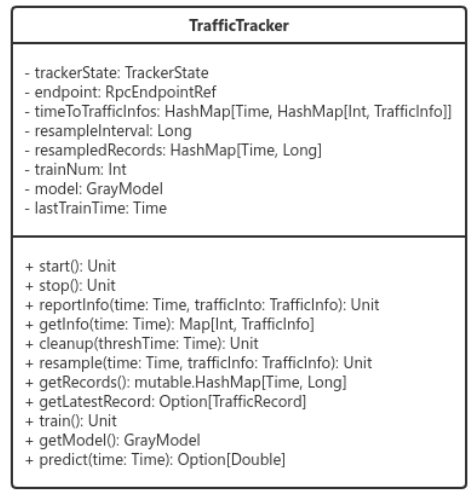

**Figure 8.** Traffictracker class diagram

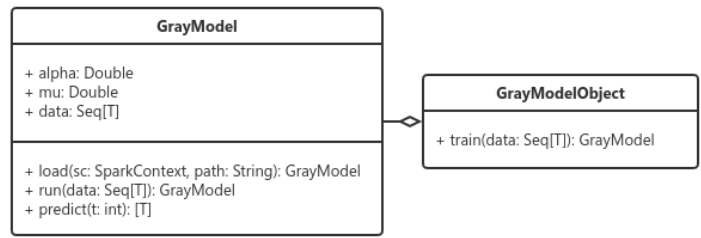

**Figure 9.** GrayModel class diagram

Figure 9 shows the class diagram of GrayModel, similar to other models in Spark MLlib. We choose the way of accompanying objects in Scala to implement the gray model. GrayModelObject is an accompanying object, which provides a *train()* interface. Users can import data to be trained and trained to get an instance of GrayModel class. GrayModel as the concrete implementation class of gray model. It provides the load function *load()* to load stored model from disk, the training function *train()*, and the prediction function *predict()* for the future data.

## 4.3 Fuzzy Control Based Batch Interval Dynamic Adjustment Layer

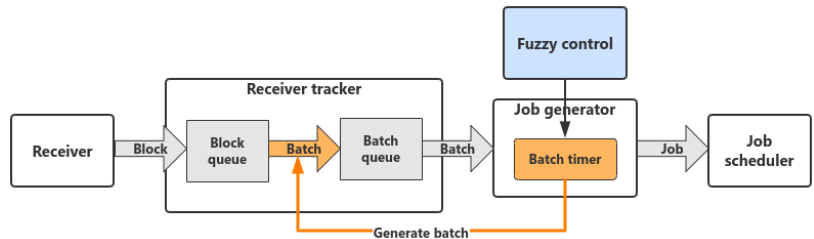

**Figure 10.** Spark streaming data division process

Figure 10 shows the data division process in Spark Streaming. Receiver periodically encapsulates the received data as Block submitted to Receiver Tracker. Receiver Tracker records all generated Block information, and adds the newly received Block to Block Queue. When receiving the signal from the Batch Timer, the Receiver Tracker will encapsulate all the Block in Block Queue into Batch, enter Batch Queue, wait for Job Generator to generate the corresponding Job, and submit it to Job Scheduler for scheduling execution.

In the vanilla version of Spark Streaming, the Batch Timer in Job Generator is an instantiated RecurringTimer class. It only supports fixed cycle triggering. In order to meet the needs of dynamic adjustment of Batch Interval, we have implemented a periodically variable timer DynamicTimer and adjusted it through Fuzzy Control module.

The Fuzzy Control module receives two inputs as the basis for adjusting Batch Interval. Among them, the prediction of vehicle flow change comes from the TrafficTracker of the traffic monitor and prediction layer, and the monitoring of the system workload comes from the StatsTracker module. Figure 11 shows the class diagram of StatsTracker, which inherits from StreamingListener. By implementing the two interfaces of *onBatchSubmitted()* and *onBatchCompleted()* provided by the StreamingListener, we can monitor the running state of Batch. When Batch is executed, StatsTracker extracts the total delay as *totalDelay* from the received BatchInfo, and divides it from Batch Interval to get a sampling of the system workload. An exponential smoothing method is then used to update the estimated values of the system workload.

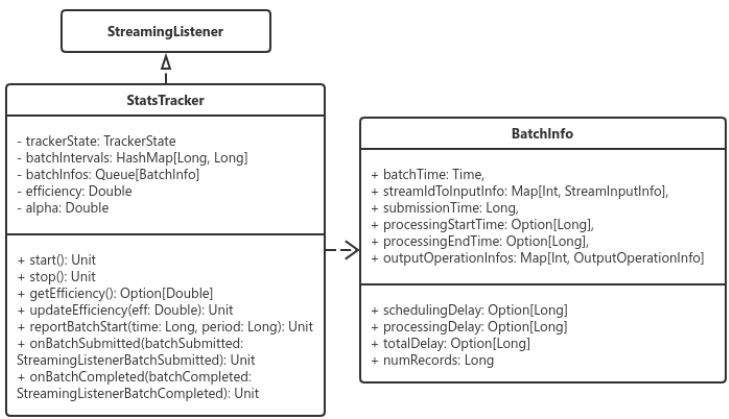

**Figure 11.** StatsTracker class diagram

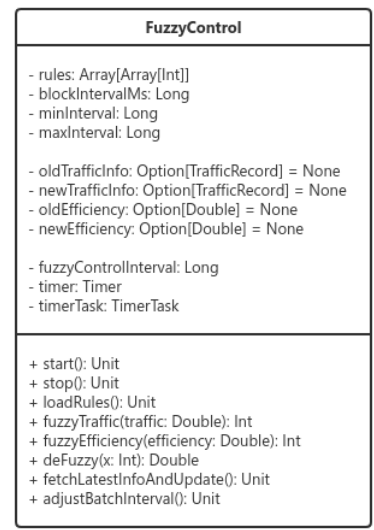

**Figure 12.** FuzzyControl class diagram

Fuzzy Control module is the core module to realize dynamic adjustment of Batch Interval. It is responsible for loading fuzzy control rules, adjusting Batch Interval according to traffic flow information and system workload information periodically. Figure 12 shows Fuzzy Control class diagram, which contains a number of attributes and methods, *rules* records the specific rules of fuzzy control, *blockIntervalMs* records the size of Block Interval. When adjusting Batch Interval, it should be adjusted by the integer multiple of Block Interval. *minInterval* and *maxInterval* limit the maximum and minimum of Batch Interval adjustment. *oldTrafficInfo* and *newTrafficInfo* record the most recent extracted traffic volume and predicted new vehicle traffic. *oldEfficiency* and *newEfficiency* record the recent changes in system workload. *fuzzyControllInterval* is the regulation cycle of fuzzy control. *timer* and *timerTask* are timers and cyclical control tasks respectively. *start()* and *stop()* methods are responsible for starting and stopping control modules. *loadRules()* method is responsible for loading fuzzy control rules. The *fuzzyTraffic()* and *fuzzyEfficiency()* methods are responsible for fuzzing the change of vehicle flow and the system workload value. The *deFuzzy()* method is responsible for defuzzification of the results obtained according to the fuzzy control rule table. *fetchLatestInfoAndUpdate()* is responsible for updating the latest vehicle flow and system workload information before every paste control starts. *adjustBatchInterval()* is the core method of this module, which is triggered by the timer periodically. It is responsible for completing data update, fuzzification, finding fuzzy control rules table, defuzzification, triggering Batch Interval adjustment and so on.

## 5. EXPERIMENTAL EVALUATION

### 5.1 Experiment Setup

In this section, we implement and test our proposed edge-oriented streaming data processing framework. Our experimental configurations are as follows:

- CPU: Intel(R) Xeon(R) CPU e5-2609 v4@1.70ghz, dual processor
- Memory: 32GB
- Disk: 4TB
- OS: Ubuntu 18.04 LTS
- Software: Spark 2.3.0, JDK 1.8, Scala 2.11.12.

### 5.2 Gray Model Based Traffic Flow Monitor and Prediction Layer Verification

In order to validate the traffic flow monitor based on the gray model and the fitting effect of the prediction layer on the vehicle flow, we design and implement two methods: the offline verification method and the online verification method.

The offline verification method is mainly used to verify the fitting effect of the gray model on the real traffic flow. We utilize the online taxi-hailing dataset provided by Didi Chuxing, which includes the whole online taxi-hailing trip records in a certain area during the period from October to November in 2016 (Didi et al., 2020). We resample the dataset with the sampling interval of 10 minutes and extract the count of drivers as the traffic flow data. Figure 13 shows the result of traffic data fitting for one of these days based on the gray model, whose horizontal axis and vertical axis relatively represent the time and the traffic flow. *y_origin* is for the original traffic flow, *y_predict* is the fitting result for traffic data by using GM, and *abs_error* stands for the absolute error between the formers. We can see that the fitting result of the gray model and the real traffic curve is very close. Figure 14 shows the percentage of relative error between *y_origin* and *y_predict*, and the average percentage of relative error in this day is about 4%. The maximum relative error appears near the 20th sampling point, which is about 18%, and the corresponding data shows the minimum traffic of this day occurs at this point so that the error is relatively increased. For offline verification, we can conclude that the gray model can better predict the short-term traffic flow, and the error is within the acceptable range.

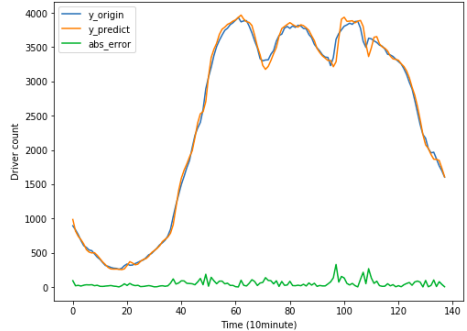

**Figure 13.** The fitting result of GM to traffic flow in one day

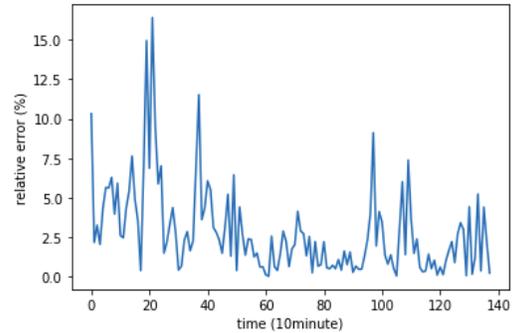

**Figure 14.** The error percentage of GM fitting

The online verification method is mainly to verify: (1) whether the traffic flow data extracted by the traffic monitor module is correct and (2) whether the traffic flow prediction module based on GM can correctly conduct the real-time prediction of short-term traffic flow. To implement the online validation, we first implement a streaming data generator to simulate data rates over time. We generate a data stream with the data arrival rate following the sinusoidal variation as shown in Figure 15. Each point in the figure represents the amount of data generated within 1 second. We use the traffic monitor module to monitor the data rate received with the sampling cycle of 30 seconds. Then, we calculate the average data rate per cycle, and use the gray model for fitting. Figure 16 shows the result of the monitoring and fitting, where *y_real* represents the real data rate observed by the Traffic Tracker module, *y_predict* refers to the real-time data rate predicted by GM, and *abs_error* represents the absolute error between them. Our observation is that the fitting error is small, which conforms to the monitor result of the data generator. By calculation, the average percentage of error between *y_real* and *y_predict* is 2.9%, and the maximum percentage is 6.4%.

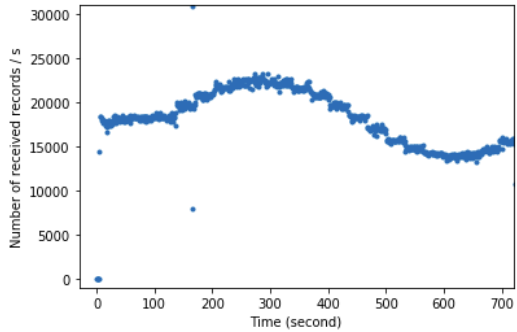

**Figure 15.** Data rate generated by the Data Generator

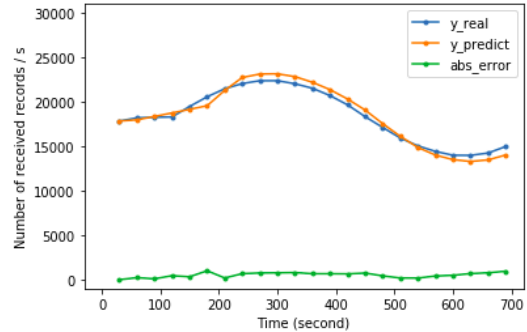

**Figure 16.** Data rate monitored by the Traffic Tracker and the fitting result of GM

## *5.3 Dynamic Division of Batch Interval Based on Fuzzy Control*

In order to verify the performance of the dynamic division of batch interval based on fuzzy control, we designed three experiments to test it comprehensively.

Experiment 1 tests the convergence performance of batch interval when the data rate is constant. We keep the data arrival rate unchanged as shown in Figure 17, and then make the batch interval slightly greater than the optimal value to observe its convergence. Figure 18 shows the change of batch interval under the fuzzy control regulation. It can be seen that its initial value is 2000ms and the fuzzy control starts after 30s. After that, the batch interval drops rapidly and converges to 1600ms. Figure 20 shows the change of system workload during this period. It can be seen that the workload of the sampled system is low at the beginning because the batch interval is much larger than the total delay. With the adjustment of the batch interval, the system workload rises rapidly, and finally stabilizes to about 0.95, which shows that the system has reached a stable state. The state can not only meet the throughput requirements of the application, but also ensure the minimum end-to-end delay.

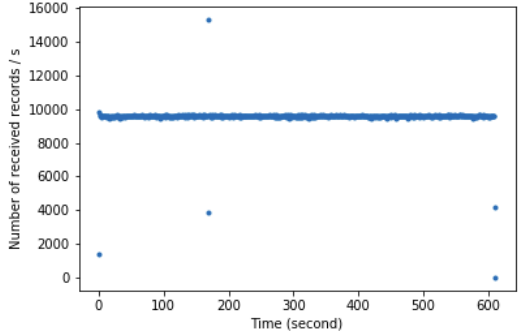

Figure 17. Data arrive rate in experiment 1

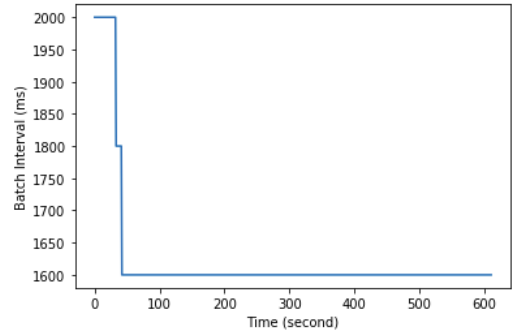

Figure 18. Batch Interval in experiment 1

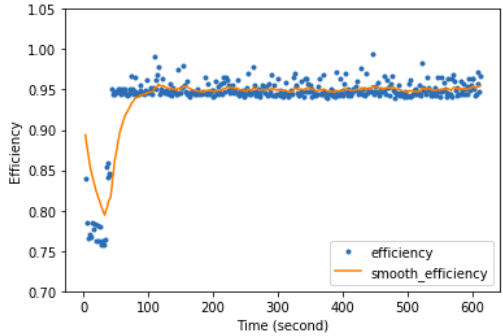

Figure 19 System workload in experiment 1

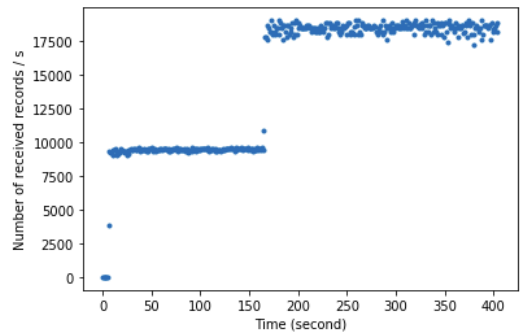

Figure 20. Data arrive rate in experiment 2

In Experiment 2, the convergence performance of batch interval is tested when the data rate changes abruptly. We set the initial batch interval of the system to a slightly greater value than the optimal value. After running for a period of time, the data arrival rate doubles and the initial batch interval of the system is less than the optimal value. Figure 20 shows the change of data arrival, from which we can see that the data arrival rate of the system has doubled after 150s. Figure 20 shows the change of batch interval. The initial value is 2000ms, and then it gradually decreases. After 50s, it converges to 1600ms. After 180s, the response data rate of fuzzy control changes abruptly. The batch interval starts to adjust and converges to 3400ms at 300s. Figure 22 shows the change of system workload. It can be seen that the system workload is lower in the first 50s because batch interval is significantly larger than total delay. After the fuzzy control is started, the batch interval is reduced, and the resource utilization rate is improved. During 70s to 150s, the system workload converges to about 0.9. After 150s, due to the sudden increase of the input data rate, the processing capacity of the system did not keep up with it in time, resulting in the total delay greater than the batch interval and the system workload greater than 1. The fuzzy control quickly responds to this change, continuously extending the batch interval to reduce the system workload, as shown in the curve of 170s to 300s. After 300s, the batch interval is stable, and the system workload converges to about 0.9.

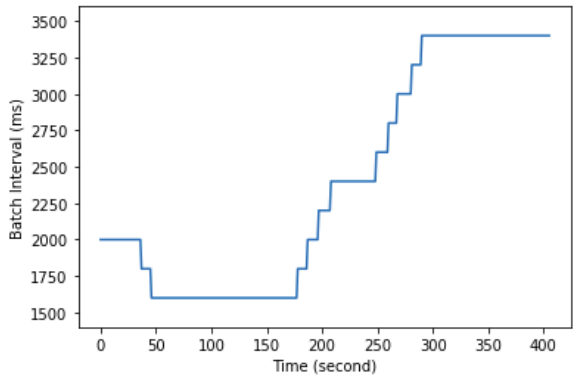 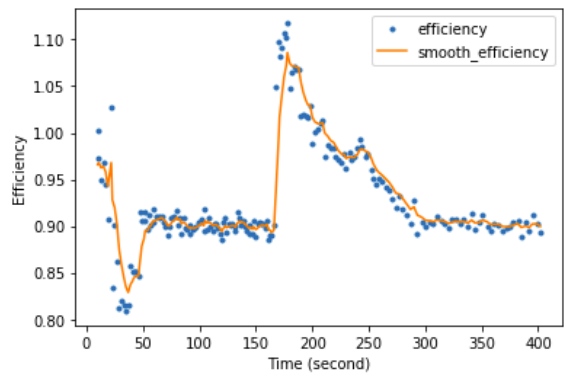

**Figure 21.** Batch Interval in experiment 2    **Figure 22.** System workload in experiment 2

Experiment 3 tests the tracking ability of the batch interval when the data rate is changing. We use the traffic flow data with sinusoidal characteristics as the input, as shown in Figure 16. Figure 23 shows the change of the batch interval and it can be seen that it changes with the change of input data rate and basically has sinusoidal characteristics. In order to test the role of the traffic flow prediction module here, Figure 25=4 shows the effect of closing the traffic forecast on the adjustment of the batch interval. Through comparison, it can be found that the introduction of traffic flow prediction makes the adjustment of the batch interval more active, while the difference between adjacent batch intervals is also reduced, avoiding the large fluctuation of the batch interval. Figure 25 shows the change of the system workload. It can be seen that although the system workload fluctuates due to the change of input data rate, the proposed framework can reduce the system workload rapidly when the system workload is greater than 1, thus ensuring the stability of the system. Figure 26 shows the processing delay and total delay of the batch, which is helpful further analyze the impact on system stability when the data arrival rate temporarily exceeds the processing capacity of the system. When the processing delay is close to the total delay, it indicates that there is little task accumulation in the system. However, when the total delay is far greater than the processing delay, the phenomenon of task accumulation is serious. These results prove that our framework can effectively inhibit the growth of total delay and ensure the stability of the system.

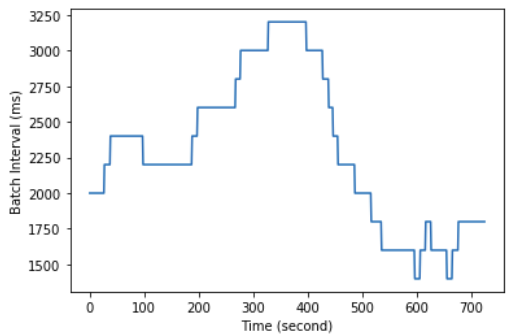 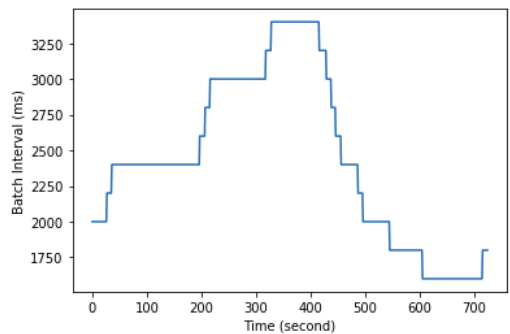

**Figure 23.** Batch Interval in experiment 3    **Figure 24.** Batch Interval in experiment 3
(disable traffic prediction)

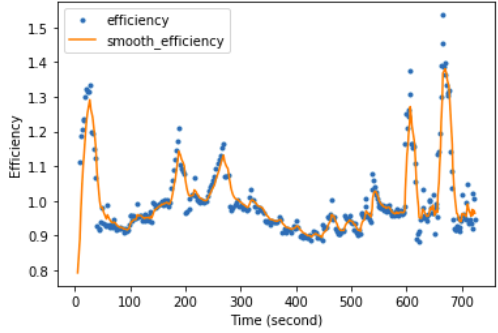
**Figure 25.** System workload in experiment 3

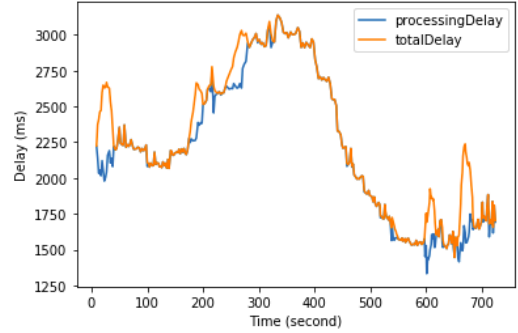
**Figure 26.** Process Delay and Total Delay in experiment 3

## 5.5 Experiment in an actual application scenario

In order to verify the effect of our proposed framework in the actual scenario, we test it based on the real traffic flow dataset. First, we choose the taxi-hailing traffic data from Didi Chuxing and extract the data from 8:00 to 20:00 on October 1, 2016. As shown in Figure 27, traffic in the morning presents an increasing trend, whereas it shows a trough around 13:00 because of the lunch break. Then, traffic recovers in the afternoon until another trough about 17:00. At 18:00 evening peak appears, after which the trend falls. We use this traffic flow as the input data rate of the experiment, which truly reflects the performance of our proposed framework in the real environment.

First, we examine the processing using the original Spark Streaming. We set the batch interval as 4s. Figure 28 shows the change in its processing delay. Shortly after the program beginning, due to the increase of data arrival rate, the processing delay keeps rising higher than 4s and stays above 4s ever since. Figure 29 presents a variation diagram of the end-to-end delay, which is increasing because the processing delay is bigger than the batch interval after running for a while, leading to the task accumulation. Figure 30 shows the change of the system workload, which is consistent with the end-to-end delay, increasing and finally reaching up to 27. This phenomenon indicates that every time a new batch of data arrives, original Spark Streaming must wait for the previous 26 batches of data to be processed first. If the task goes on for a long time, the system easily gets crashed due to insufficient memory.

Next, we test the processing performance of the implemented framework. Figure 31 shows the variation of the end-to-end delay under fuzzy control. We can see that the end-to-end delay changes as the data arrival rate changes, but no significant task accumulation occurs. Figure 32 shows the change of the batch interval. In order to adapt to the change of input data rate, the system workload is constantly adjusted to be in the best state. The change of system workload is shown as Figure 34, from which we can observe that our processing framework keeps the system workload around 0.95 by constant adjustment.

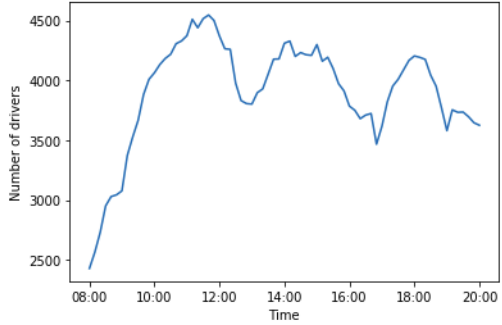

**Figure 27.** Traffic flow information for a day in reality

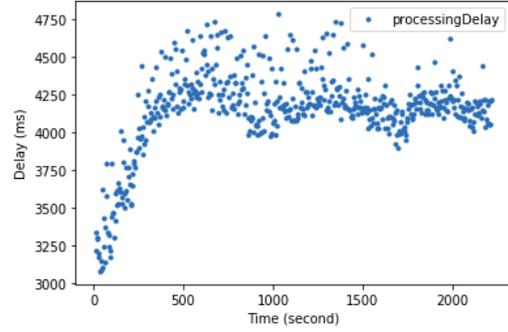

**Figure 28.** Processing Delay in vanilla version Spark Streaming

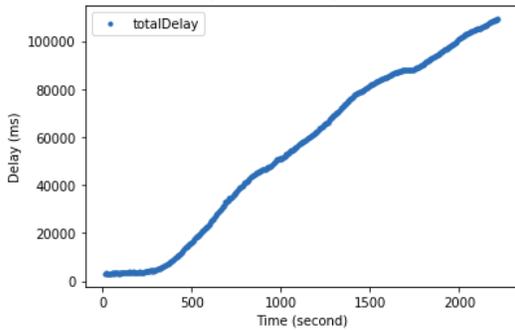

**Figure 29.** Total Delay in vanilla version Spark Streaming

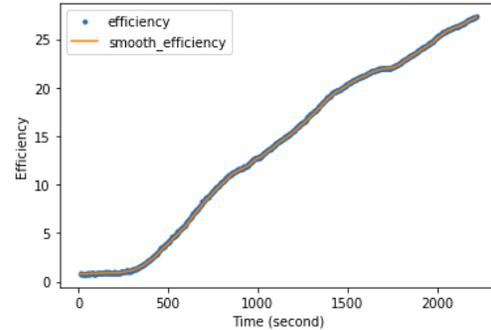

**Figure 30.** System workload in vanilla version Spark Streaming

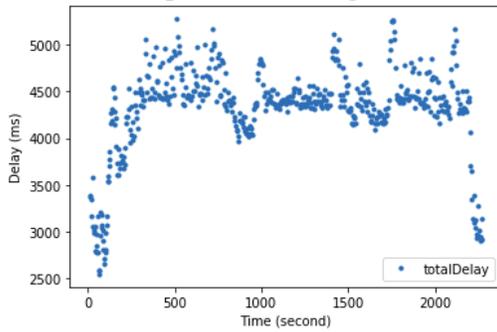

**Figure 31.** Total Delay when using fuzzy control

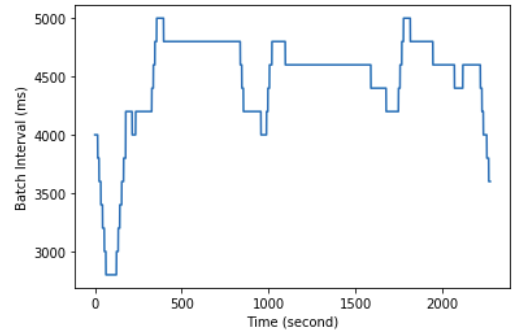

**Figure 32.** Batch Interval

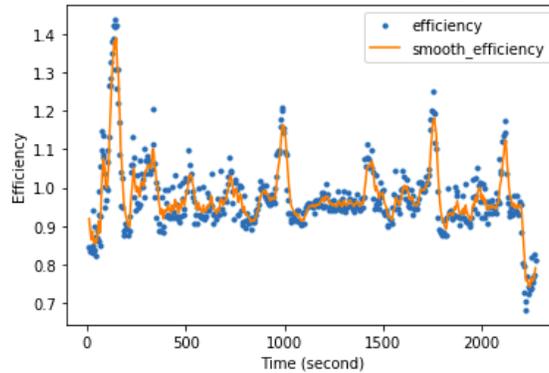

**Figure 33.** System workload

# 6. RELATED WORK

- Autonomous Driving

Autonomous driving is an attractive research field with the development of sensor technology and deep learning, with the L2 and L3 conditional driving automation technology become matured, many companies and researches divert attention to L4 even L5 full driving automation. An ideal autonomous driving can reduce lots of traffic accidents due to human mistakes, such as drowsiness (Kholerdi, H. A., et al., 2015). Liu, S. et al. shared their practical experiences of creating autonomous vehicle system, discussed traditional techniques and deep learning based techniques used for autonomous driving (Liu, S. et al., 2017). Autonomous vehicle system is very complex, has a lot of sensors data process and machine learning tasks, which always inadvertently beyond on-board computer process power, Liu S. et al. proposed a unified cloud platform to enhanced performance and energy efficiency of autonomous driving application in cloud platform (Liu, S. et al., 2017). de Cock Buning, M. et al. discussed the moral and legal questions about autonomous driving (de Cock Buning, M., et al., 2017).

- Stream Processing Frameworks

Along with the increase in Streaming data, and the increasing demand for its processing, a large number of streaming data processing frameworks have emerged. According to their different data divisions, they can be divided into continuous stream processing frameworks and micro-batch stream processing frameworks. Yahoo! S4 (Neumeyer, L. et al., 2010), Storm (Toshniwal, Taneja, Shukla et al., 2017) and Apache Samza (Noghabi, Paramasivam, Pan et al., 2017) is the representative of continuous stream processing framework, and Spark Streaming is the representative of micro-batch stream processing frameworks. Among them, Yahoo! S4 provided a event based distributed streaming processing platform, which could quickly response to high frequency events, but no guarantee for reliable processing, events maybe losed when system overload. Storm use data flow model, which could make sure streaming data been reliable processing, while throughput is relatively low compared to Spark Streaming. Apache Samze is a streaming processing framework based on Apache Kafka, and it has a callback-based process message API. Spark Streaming is a streaming processing framework above Spark, and it use DAG to describe execute flow, received data been divided to some batch then send to Spark for execution. In addition, Bhatotia P et al proposed Incoop, an incremental data processing framework, to allow users to reuse calculation results as much as possible (Bhatotia, Wieder, Rodrigues et al., 2011).

- Traffic Prediction and Control

The emergence of the vehicle network, makes road traffic flow with streaming data traffic become strongly association. Lai. Y. et al. analyze the trajectory data sets of urban taxis, conclude urban traffic Coulomb's law, and based on it to proposed a route recommendation scheme (Lai, Y. et al., 2018). Researchers have used many methods to predict road traffic flow in short-term or long-term and achieve high accuracy, such as Markov chain (Yu, G., 2003), gray model (XU, C., et al., 2010), and LSTM (Fu, R., et al., 2016), etc. In order to cope with the fluctuation of streaming data, (Amini, Jain et al., 2006; Schneider, Andrade et al., 2009) dynamically changed system resources to provide sufficient flexibility, but in most cases the system resources are limited. Babcock B, et al propose the workload shedding in response to burst traffic, but when the traffic is too large, some data will be discarded (Babcock, Datar et al., 2004). Spark Steaming introduced the Back Pressure mechanism in version 1.5 to limit the input traffic.

- Spark and Spark Streaming optimization

As a general-purpose big data processing framework, Spark provides a rich set of parameters to suit the needs of different deployment environments, which have been extensively optimized by researchers either manually or automatically. Wang G et al used machine learning technology to automatically optimize the parameters in Spark (Wang, Taneja et al., 2016). In Spark deploy practice, memory resource insufficient is always a bottleneck to limit in-memory big data computing performance, Wang, B. et al. proposed a hybrid DRAM/NVM memory architecture to solve the memory capacity and energy consumption dilemmas for in-memory big data computing systems (Wang, B. et al., 2019)(Wang, B. et al., 2019). Das T et al found that the processing time of different operators in Spark Streaming varies with the data rate, and proposed a Batch Interval adjustment algorithm using fixed-point iteration (Das, Zhong et al., 2014). Zhang Q et al proposed DyBBS and realized joint optimization of block interval and batch interval, which can better reduce end-to-end delay (Zhang, Song, Routray et al., 2016). Liang Y et al proposedWang a parameter and resource coordination adjust strategy, using dynamic neighborhood particle swarm algorithm to find a resource minimization solution (Liang, Liu, Routray et al., 2019). Cheng D et al proposed A-scheduler, which divided Spark Streaming tasks into data-independent tasks and data-dependent tasks, and scheduled them respectively to improve the concurrency and resource utilization efficiency of the system and realized the adjustment of Batch Interval based on the fuzzy control method of the expert system. (Wang et al., 2018).

## 7. CONCLUSION

Wide deployment autonomous driver sensors generate a large amount of stream data all the time, traditional cloud computing can't meet the latency requirements. To deal with this problem, considering the difference between cloud computing and edge computing, we proposed a stream processing framework for edge computing by migrating stream data processing from cloud datacenter into edge datacenter.

Considering the volatility character of streaming data arrival rate, our proposed framework implements gray model based traffic flow monitor and prediction, so that it could forecast data flow changes and make responds in advance. Experiments show that the gray model based traffic flow monitor prediction could fit short-term traffic very well.

Considering the latency sensitive character of autonomous driving applications, our proposed framework implements fuzzy control based Batch Interval adjustment mechanism, so that it could adjust Batch Interval according to running environment. Experiments show that the Batch Interval adjustment mechanism could quickly respond and decline end-to-end latency as much as possible meanwhile do not damage the throughput. Experiments also show that our proposed framework could effectively guarantee the stability of system when system workload changes.